\documentclass{llncs}
\usepackage[utf8]{inputenc}
\usepackage{epsfig}
\usepackage{multicol}
\usepackage{multirow}
\usepackage{wrapfig}
\usepackage{fancyvrb}
\usepackage{url}

\begin{document}

\title{Efficient Support for Mode-Directed Tabling in the YapTab Tabling System}

\author{João Santos \and Ricardo Rocha}

\institute{CRACS \& INESC TEC, Faculty of Sciences, University of Porto\\
           Rua do Campo Alegre, 1021/1055, 4169-007 Porto, Portugal\\
           \email{\{jsantos,ricroc\}@dcc.fc.up.pt}}

\maketitle


\begin{abstract}
  Mode-directed tabling is an extension to the tabling technique that
  supports the definition of mode operators for specifying how answers
  are inserted into the table space. In this paper, we focus our
  discussion on the efficient support for mode directed-tabling in the
  YapTab tabling system. We discuss 7 different mode operators and
  explain how we have extended and optimized YapTab's table space
  organization to support them. Initial experimental results show that
  our implementation compares favorably with the B-Prolog and XSB
  state-of-the-art Prolog tabling systems.
\end{abstract}


\section{Introduction}

Tabling~\cite{Chen-96} is a recognized and powerful implementation
technique that solves some limitations of Prolog's operational
semantics in dealing with recursion and redundant
sub-computations. Tabling based models are able to reduce the search
space, avoid looping, and always terminate for programs with the
\emph{bounded term-size property}. Tabling consists of saving and
reusing the results of sub-computations during the execution of a
program and, for that, the calls and the answers to tabled subgoals
are stored in a proper data structure called the \emph{table
  space}. In a traditional tabling system, all the arguments of a
tabled subgoal call are considered when storing answers into the table
space. When a new answer is not a variant\footnote{Two (answer or
  subgoal) terms are considered to be variant if they are the same up
  to variable renaming.} of any answer that is already in the table
space, then it is always considered for insertion. Therefore,
traditional tabling systems are very good for problems that require
storing all answers. \emph{Mode-directed tabling}~\cite{Guo-08} is an
extension to the tabling technique that supports the definition of
selective criteria for specifying how answers are inserted into the
table space. The idea of mode-directed tabling is to use \emph{mode
  operators} to define what arguments should be used in variant
checking in order to select what answers should be tabled.

In a traditional tabling system, to evaluate a predicate $p/n$ using
tabling, we just need to declare it as `$table~p/n$'. With
mode-directed tabling, tabled predicates are declared using statements
of the form `$table~p(m_1,...,m_n)$', where the $m_i$’s are mode
operators for the arguments. Implementations of mode-directed tabling
are already available in systems like ALS-Prolog~\cite{Guo-08} and
B-Prolog~\cite{Zhou-10}, and a restricted form of mode-directed
tabling can be also recreated in XSB Prolog by using \emph{answer
  subsumption}~\cite{Swift-10}.

In this paper, we focus our discussion on the efficient implementation
of mode directed-tabling in the YapTab tabling
system~\cite{Rocha-05a}, which uses
\emph{tries}~\cite{RamakrishnanIV-99} to implement the table
space. Our implementation uses a more general approach to the
declaration and use of mode operators and, currently, it supports 7
different modes: \emph{index}, \emph{first}, \emph{last}, \emph{min},
\emph{max}, \emph{sum} and \emph{all}. To the best of our knowledge,
no other tabling system supports all these modes and, in particular,
the \emph{sum} mode is not supported by any other system. Experimental
results, using a set of benchmarks that take advantage of
mode-directed tabling, show that our implementation compares favorably
with the B-Prolog and XSB state-of-the-art Prolog tabling systems.

The remainder of the paper is organized as follows. First, we
introduce some background concepts about tabling. Next, we describe
the mode operators that we propose and we show some small examples of
their use. Then, we introduce YapTab's table space organization and
describe how we have extended it to efficiently support mode-directed
tabling. At last, we present some experimental results and we end by
outlining some conclusions.


\section{Tabled Evaluation}

In a traditional tabling system, programs are evaluated by storing
answers for tabled subgoals in an appropriate data structure called
the \emph{table space}. Similar calls to tabled subgoals are not
re-evaluated against the program clauses, instead they are resolved by
consuming the answers already stored in the corresponding table
entries. During this process, as further new answers are found, they
are stored in their tables and later returned to all similar calls.

\begin{figure}[t]
\centering
\includegraphics[width=10.5cm]{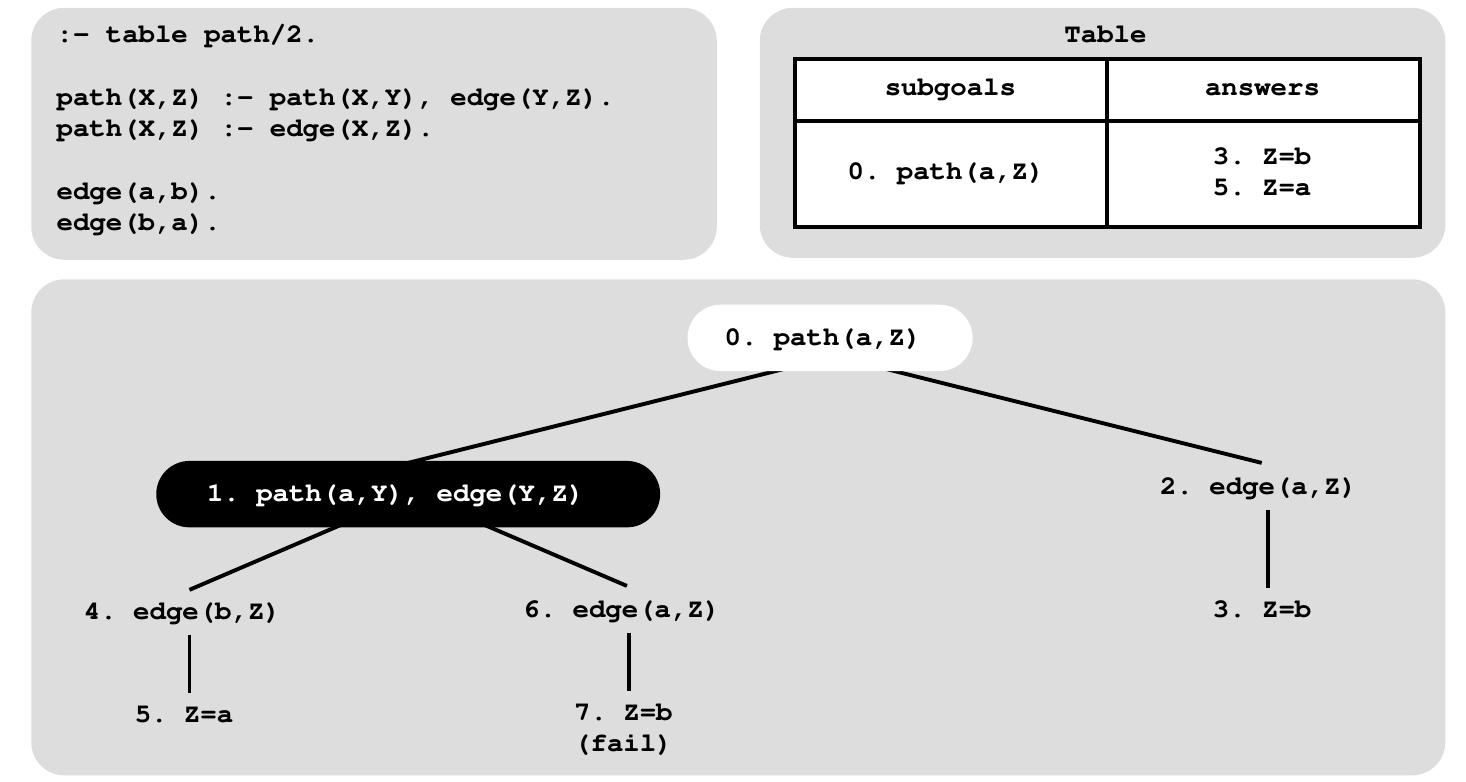}
\caption{An example of a tabled evaluation}
\label{fig_tabled_eval}
\vspace{-\bigskipamount}
\end{figure}

Figure~\ref{fig_tabled_eval} illustrates the execution of a tabled
program. The top left corner of the figure shows the program code and
the top right corner shows the final state of the table space. The
program defines a small directed graph, represented by two
\emph{edge/2} facts, with a relation of reachability, defined by a
\emph{path/2} tabled predicate. The bottom of the figure shows the
evaluation sequence for the query goal \emph{path(a,Z)}. Note that
traditional Prolog would immediately enter an infinite loop because
the first clause of \emph{path/2} leads to a variant call to
\emph{path(a,Z)}.

First calls to tabled subgoals correspond to generator nodes (nodes
depicted by white oval boxes) and, for first calls, a new entry,
representing the subgoal, is added to the table space (step 0). Next,
\emph{path(a,Z)} is resolved against the first matching clause
calling, in the continuation, \emph{path(a,Y)} (step 1). Since
\emph{path(a,Y)} is a variant call to \emph{path(a,Z)}, we do not
evaluate the subgoal against the program clauses, instead we consume
answers from the table space. Such nodes are called \emph{consumer
  nodes} (nodes depicted by black oval boxes). However, at this point,
the table does not have answers for this call, so the computation is
suspended.

The only possible move after suspending is to backtrack and try the
second matching clause for \emph{path(a,Z)} (step 2). This originates
the answer \{\emph{Z=b}\}, which is then stored in the table space
(step 3). At this point, the computation at node 1 can be resumed with
the newly found answer (step 4), giving rise to one more answer,
\{\emph{Z=a}\} (step 5). This second answer is then also inserted in
the table space and propagated to the consumer node (step 6), which
originates the answer \{\emph{Z=b}\} (step 7). This answer had already
been found at step 3. Tabling does not store duplicate answers in the
table space and, instead, repeated answers \emph{fail}. This is how
tabling avoids unnecessary computations, and even looping in some
cases. A new answer is inserted in table space only if it is not a
variant of any answer that is already there. Since there are no more
answers to consume nor more clauses left to try, the evaluation ends
and the table entry for \emph{path(a,Z)} can be marked as
\emph{completed}.


\section{Mode-Directed Tabling}

With mode-directed tabling, tabled predicates are declared using
statements of the form `$table~p(m_1,...,m_n)$', where the $m_i$’s are
\emph{mode operators} for the arguments. We have defined 7 different
mode operators: \emph{index}, \emph{first}, \emph{last}, \emph{min},
\emph{max}, \emph{sum} and \emph{all}. Arguments with modes
\emph{first}, \emph{last}, \emph{min}, \emph{max}, \emph{sum} or
\emph{all} are assumed to be output arguments and only \emph{index}
arguments are considered for variant checking. After an answer be
generated, the system tables the answer only if it is
\emph{preferable}, accordingly to the meaning of the output arguments,
than some existing variant answer. Next, we describe in more detail
how these modes work and we show some examples of their use in the
YapTab system.


\subsection{Index/First/Last Mode Operators}

\begin{figure}[t]
\centering
\includegraphics[width=12cm]{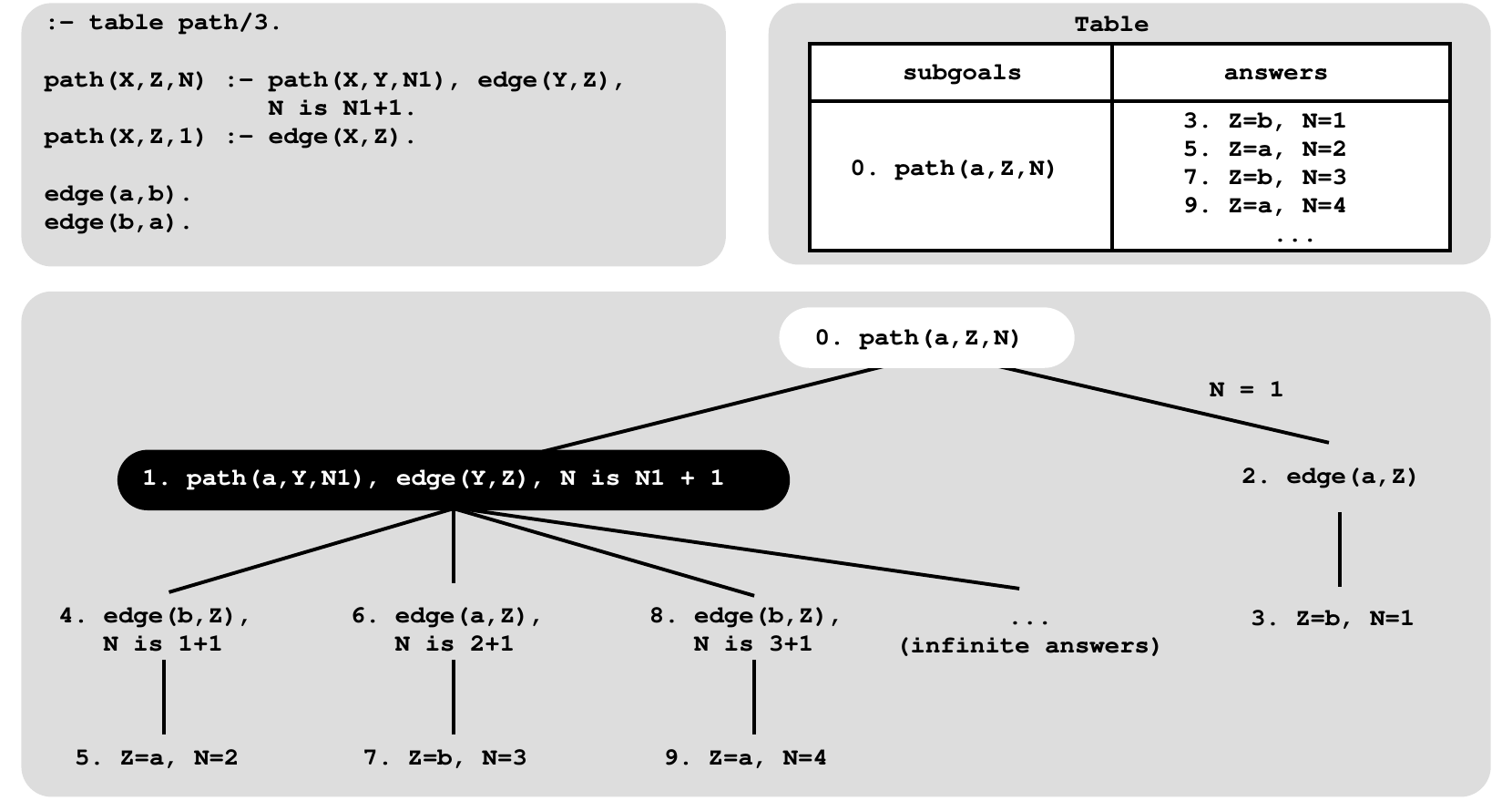}
\caption{A tabled evaluation with an infinite number of answers}
\label{fig_tabled_eval_inf}
\vspace{-\bigskipamount}
\end{figure}

Starting from the example in Fig.~\ref{fig_tabled_eval}, consider now
that we modify the program so that it also calculates the number of
edges that are traversed in a path. Figure~\ref{fig_tabled_eval_inf}
illustrates the execution of this new program. As we can see, even
with tabling, the program does not terminates. Such behavior occurs
because there is a path with an infinite number of edges starting from
\emph{a}, thus not verifying the bounded term-size property necessary
to ensure termination. In particular, the answers found at steps 3 and
7 and at steps 5 and 9 have the same answer for variable \emph{Z}
(\{\emph{Z=b}\} and \{\emph{Z=a}\}, respectively), but they are both
inserted in the table space because they are not variants for variable
\emph{N}.

Knowing that the problem with the program in
Fig.~\ref{fig_tabled_eval_inf} resides on the fact that the third
argument generates an infinite number of answers, we can thus define
the \emph{path/3} predicate to have mode
\emph{path(index,index,first)}. The \emph{index} mode means that only
the given arguments must be considered for variant checking. The
\emph{first} mode means that only the first answer must be stored. By
considering this declaration, the answer \{\emph{Z=b,~N=3}\} is no
longer inserted in the table and execution fails. That happens
because, with the \emph{first} mode on the third argument, the answer
\{\emph{Z=b,~N=1}\} found at step 3 is considered a variant of the
answer \{\emph{Z=b,~N=3}\} found at step 7.

The \emph{last} mode implements the opposite behavior of the
\emph{first} mode, i.e., it always stores the last answer being found
and deletes the previous one, if any. The \emph{last} mode has shown
to be very useful for implementing problems involving
Preferences~\cite{Guo-05b} and Answer Subsumption~\cite{Santos-12}.


\subsection{Min/Max Mode Operators}

\begin{figure}[t]
\centering
\includegraphics[width=12cm]{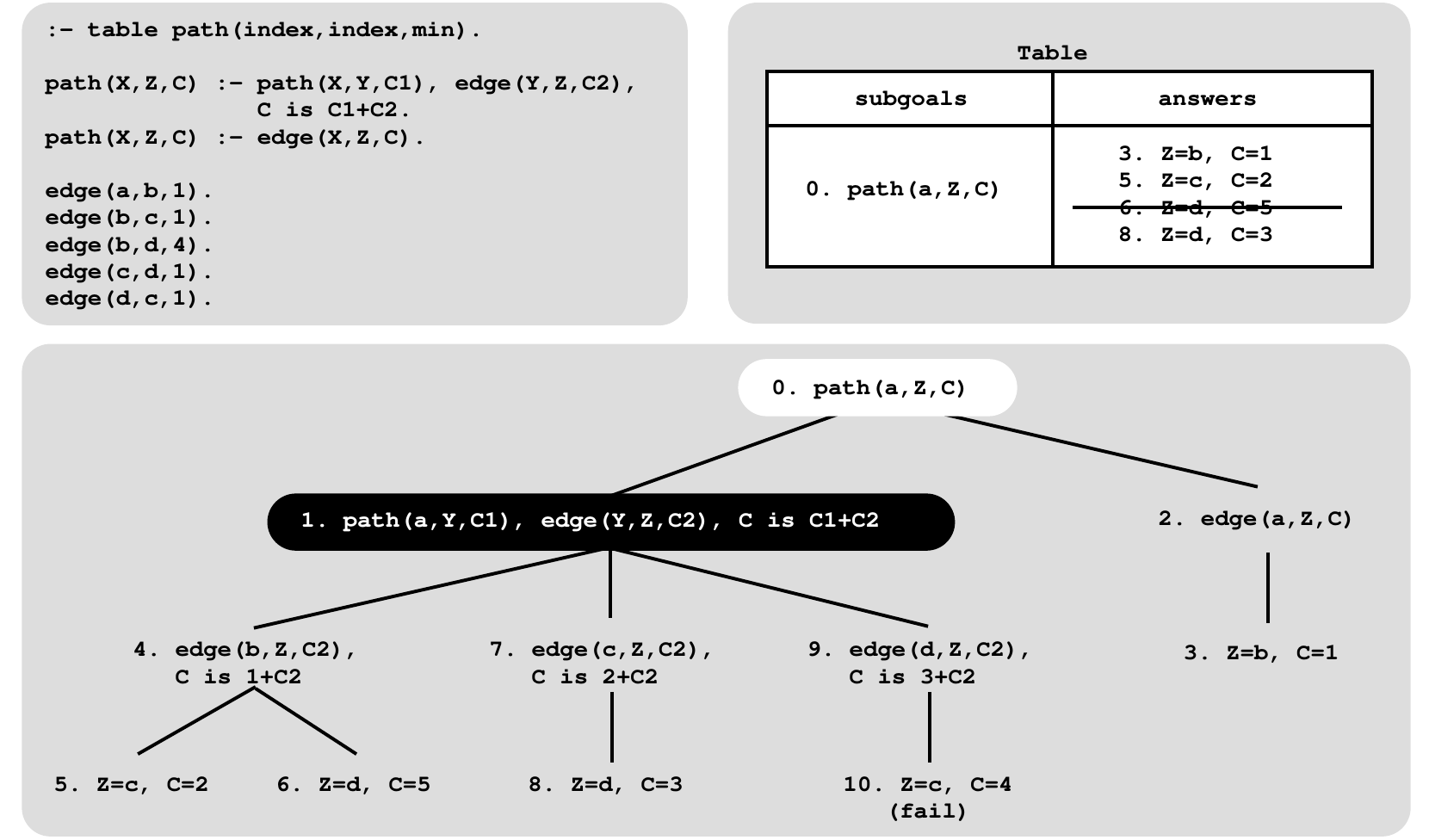}
\caption{Using the \emph{min} mode to compute the paths with the
  shortest distances}
\label{fig_tabled_eval_min}
\vspace{-\bigskipamount}
\end{figure}

The \emph{min} and \emph{max} modes allow to specify a selective
criteria that stores, respectively, the minimal and maximal answers
found for an argument. To better understand their behavior,
Fig.~\ref{fig_tabled_eval_min} shows an example using the \emph{min}
mode. The program's goal is to compute the paths with the shortest
distances. To do that, the \emph{path/3} predicate is declared as
\emph{path(index,index,min)}, meaning that the third argument should
store only the minimal answers for the first two arguments.

By observing the example in Fig.~\ref{fig_tabled_eval_min}, we can see
that the execution tree follows the normal evaluation of a tabled
program and that the answers are stored as they are found. The most
interesting part happens at step 8, where the answer
\{\emph{Z=d,~C=3}\} is found. This answer is a variant of the answer
\{\emph{Z=d,~C=5}\} found at step 6. In the previous example, with the
\emph{first} mode, the old answer would have been kept in the
table. Here, as the new answer is minimal on the third argument, the
old answer is replaced by the new answer.

The \emph{max} mode works similarly, but stores the maximal answer
instead. In any case, we must be careful when using these two modes as
they may not ensure termination for programs without the bounded
term-size property. For instance, this would be the case if, in the
example of Fig.\ref{fig_tabled_eval_min}, we used the \emph{max} mode
instead of the \emph{min} mode.


\subsection{Sum/All Mode Operators}

\begin{figure}[t]
\centering
\includegraphics[width=12cm]{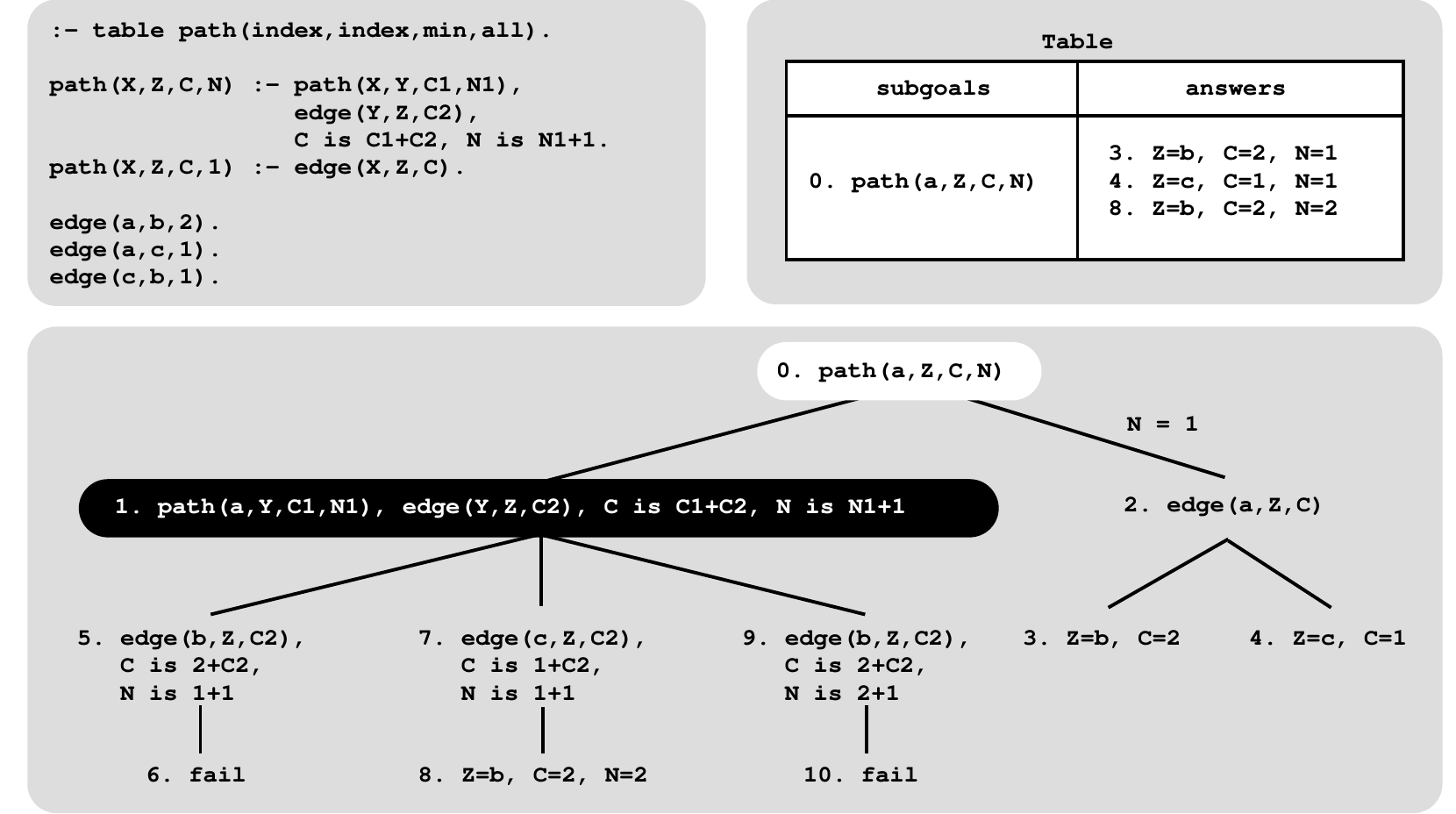}
\caption{Using the \emph{all} mode to compute the paths with the
  shortest distances together with the number of edges traversed}
\label{fig_tabled_eval_all}
\vspace{-\bigskipamount}
\end{figure}

Two other modes that can be useful are the \emph{sum} and the
\emph{all}. The \emph{sum} mode allows to sum all the answers for a
given argument and the \emph{all} mode allows to store all the answers
for a given argument. Consider, for example, the program in
Fig.~\ref{fig_tabled_eval_all} where the \emph{path/3} predicate is
declared as \emph{path(index,index,min,all)} meaning that, for each
path, we want to store the shortest distance of the path (the third
argument) and, at the same time, we want to store the number of edges
traversed, for all paths with the same minimal distances (the fourth
argument).

The execution tree for the program in Fig.~\ref{fig_tabled_eval_all}
is similar to the previous ones. The most interesting part happens
when the answer \{\emph{Z=b,~C=2,~N=2}\} is found at step 8. This
answer is a variant of the answer found at step 3 and although both
have the same minimal value (\emph{C=2}), the new answer is still
inserted in the table space since the number of edges (fourth
argument) is different.

Notice that when the \emph{sum} or \emph{all} modes are used in
conjunction with another mode, like the \emph{min} mode in the
example, it is important to keep in mind that the aggregation of
answers made for the \emph{sum} or \emph{all} argument depends on the
corresponding answer for the \emph{min} argument. Consider, for
example, that in the previous example we had found one more answer
\{\emph{Z=b,~C=1,~N=4}\}. In this case, the new answer would be
inserted and the answers \{\emph{Z=b,~C=2,~N=1}\} and
\{\emph{Z=b,~C=2,~N=2}\} would be deleted because the new answer
corresponds to a shorter distance, as defined by the value \emph{C=1}
in the \emph{min} argument.


\subsection{Related Work}
\label{sec_related_work}

The ALS-Prolog~\cite{Guo-08} and B-Prolog~\cite{Zhou-10} systems also
implement mode-directed tabling using a very similar syntax. However,
some mode operators have different names in those systems. For
example, the \emph{index}, \emph{first} and \emph{all} modes are known
as \emph{+}, \emph{-} and \emph{@}, respectively. The \emph{sum} mode
is not supported by any other system and B-Prolog also does not
implement the \emph{last} and \emph{all} modes. The \emph{+}
(\emph{index}) mode in B-Prolog is assumed to be an input argument,
which means that it can only be called with ground terms. On the other
hand, B-Prolog includes an extra mode, named \emph{nt}, to indicate
that a given argument should not be tabled and, thus, not considered
to be inserted in the table space. B-Prolog also extends the
mode-directed tabling declaration to include a \emph{cardinality
  limit} that allows to define the maximum number of answers to be
stored in the table space~\cite{Zhou-10}.

Mode-directed tabling can also be recreated in the XSB Prolog system
by using \emph{answer subsumption}~\cite{Swift-10}. XSB Prolog has two
answer subsumption mechanisms. One is called \emph{partial order
  answer subsumption} and can be used to mimic, in terms of
functionality, the \emph{min} and \emph{max} modes. Consider that we
want to use it with the program in Fig.~\ref{fig_tabled_eval_min} that
computes the paths with the shortest distances. Then, we should
declare the \emph{path/3} predicate as $path(\_,\_,po(</2))$ meaning
that the third argument will be evaluated using partial order answer
subsumption, where the predicate $</2$ implements the partial order
relation. The other two arguments are considered to be index
arguments.

The other XSB's mechanism, called \emph{lattice answer subsumption},
is more powerful and can be used to mimic, in terms of functionality,
the other modes. To use it with the same example, we only need to
change the \emph{path/3} declaration to
$path(\_,\_,lattice(min/3))$. Note that the \emph{min/3} predicate
must have three arguments. This is necessary since, with this
mechanism, we can generate a third answer starting from the new answer
and from the answer stored in the table. For example, for the shortest
path problem, the predicate \emph{min/3} could be something like:

\vspace{0.25cm}
$min(Old,New,Res):-~Old<New \rightarrow Res=Old~;~Res=New.$


\section{Implementation}

In this subsection, we describe the changes made to YapTab in order to
support mode-directed tabling. We start by briefly presenting some
background concepts about the table space organization in YapTab and
then we discuss in more detail how we have extended it to efficiently
support mode-directed tabling.


\subsection{YapTab's Table Space Organization}

Like we have seen, during the execution of a program, the table space
may be accessed in a number of ways: (i) to find out if a subgoal is
in the table and, if not, insert it; (ii) to verify whether a newly or
preferable answer is already in the table and, if not, insert it; and
(iii) to load answers from the tables.

With these requirements, a careful design of the table space is
critical to achieve an efficient implementation. YapTab uses
\emph{tries} which is regarded as a very efficient way to implement
the table space~\cite{RamakrishnanIV-99}. A trie is a tree structure
where each different path through the \emph{trie nodes} corresponds to
a term described by the tokens labeling the traversed nodes. For
example, the tokenized form of the term $path(X,1,f(Y))$ is the
sequence of 5 tokens $path/3$, $VAR_0$, $1$, $f/1$ and $VAR_1$, where
each variable is represented as a distinct $VAR_i$
constant~\cite{Bachmair-93}. Two terms with common prefixes will
branch off from each other at the first distinguishing
token. Consider, for example, a second term $path(Z,1,b)$, represented
by the sequence of 4 tokens $path/3$, $VAR_0$, $1$ and $b$. Since the
main functor, token $path/3$, and the first two arguments, tokens
$VAR_0$ and $1$, are common to both terms, only one node will be
required to fully represent this second term in the trie, thus
allowing to save three nodes in this case.

YapTab's table design implements tables using two levels of tries. The
first level, named \emph{subgoal trie}, stores the tabled subgoal
calls and the second level, named \emph{answer trie}, stores the
computed answers for a given call. More specifically, each tabled
predicate has a \emph{table entry} data structure assigned to it,
acting as the entry point for the predicate's subgoal trie. Each
different subgoal call is then represented as a unique path in the
subgoal trie, starting at the predicate's table entry and ending in a
\emph{subgoal frame} data structure, with the argument terms being
stored within the path's nodes. The subgoal frame data structure acts
as an entry point to the answer trie. Contrary to subgoal tries,
answer trie paths hold just the substitution terms for the free
variables that exist in the argument terms of the corresponding
subgoal call~\cite{RamakrishnanIV-99}.

\begin{wrapfigure}{r}{5cm}
\vspace{-\intextsep}
\centering
\includegraphics[width=5cm]{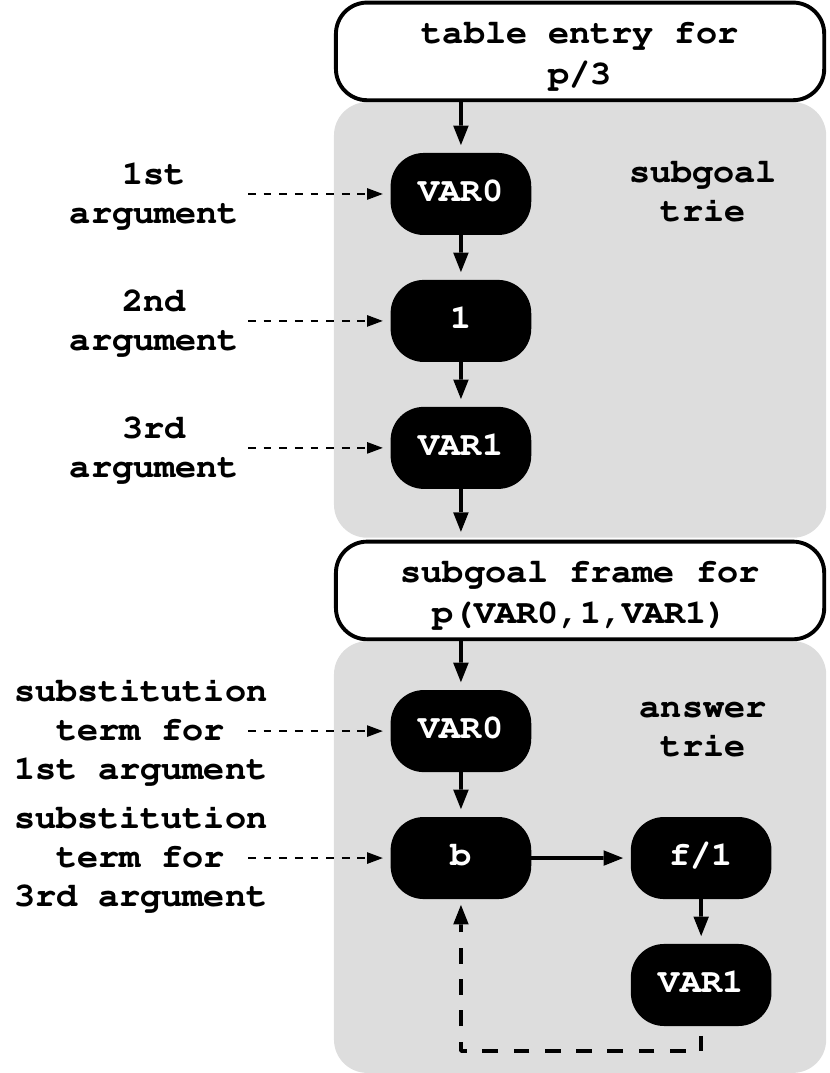}
\caption{Table space organization}
\label{fig_table_design}
\vspace{-\bigskipamount}
\end{wrapfigure}

An example for a tabled predicate $p/3$ is shown in
Fig.~\ref{fig_table_design}. Initially, the table entry for $p/3$
points to an empty subgoal trie. Then, the subgoal $p(X,1,Y)$ is
called and three trie nodes are inserted to represent the arguments in
the call: one for variable $X$ ($VAR_0$), a second for integer 1, and
a last one for variable $Y$ ($VAR_1$). Since the predicate's functor
term is already represented by its table entry, we can avoid inserting
an explicit node for $p/3$ in the subgoal trie. Then, the leaf node is
set to point to a subgoal frame, from where the answers for the call
will be stored. The example shows two answers for $p(X,1,Y)$:
\{\emph{X=$VAR_0$,~Y=f($VAR_1$)}\} and
\{\emph{X=$VAR_0$,~Y=b}\}. Since both answers have the same
substitution term for argument $X$, they share the top node in the
answer trie ($VAR_0$). For argument $Y$, each answer has a different
substitution term and, thus, a different path is used to represent
each.

When adding answers, the leaf nodes are chained in a linked list in
insertion time order, so that the recovery may happen the same way. In
Fig.~\ref{fig_table_design}, we can observe that the leaf node for the
first answer (node $VAR_1$) points (dashed arrow) to the leaf node of
the second answer (node $b$). To maintain this list, two fields in the
subgoal frame data structure point, respectively, to the first and
last answer of this list (for simplicity of illustration, these
pointers are not shown in Fig.~\ref{fig_table_design}). When consuming
answers, a consumer node only needs to keep a pointer to the leaf node
of its last loaded answer, and consumes more answers just by following
the chain. Answers are loaded by traversing the trie nodes bottom-up
(again, for simplicity of illustration, such pointers are not shown in
Fig.~\ref{fig_table_design}).


\subsection{Mode-Directed Tabled Subgoal Calls}

In YapTab, mode-directed tabled predicates are compiled by extending
the table entry data structure to include a \emph{mode array}, where
the information about the modes is stored. In this mode array, the
modes appear in the order in which the arguments are accessed, which
can be different from their position in the original declaration. For
example, \emph{index} arguments must be considered first, irrespective
of their position. Or, if using the \emph{all} and \emph{min} modes in
a declaration, all \emph{min} arguments must be considered before any
\emph{all} argument, since the \emph{all} means that all answers must
be stored, making meaningless the notion of being minimal in this
case. As we will see in
Section~\ref{sec_mode_directed_tabled_answers}, changing the order is
also strictly necessary to achieve an efficient implementation. In
YapTab, the mode information is thus stored in the order mentioned
below, together with the argument's position:

\begin{enumerate}
\item arguments with \emph{index} mode;
\item arguments with \emph{max} or \emph{min} mode;
\item arguments with \emph{all} mode;
\item argument (only one is allowed) with \emph{sum} or \emph{last} mode;
\item arguments with \emph{first} mode.
\end{enumerate}

\begin{wrapfigure}{r}{5.25cm}
\vspace{-\intextsep}
\centering
\includegraphics[width=5.25cm]{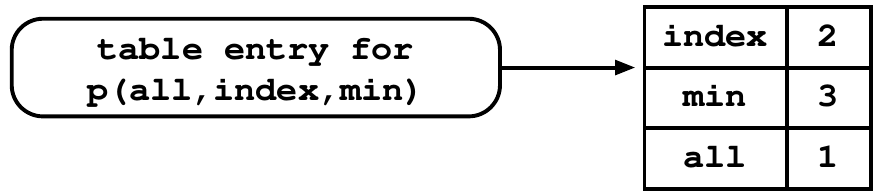}
\caption{Mode array}
\label{fig_mode_array}
\vspace{-\bigskipamount}
\end{wrapfigure}

Figure~\ref{fig_mode_array} shows an example for a
\emph{p(all,index,min)} mode-directed tabled predicate. The
\emph{index} mode is placed first in the mode array, then the
\emph{min} mode and last the \emph{all} mode.

\begin{wrapfigure}{r}{6.35cm}
\vspace{-\intextsep}
\centering
\includegraphics[width=6.35cm]{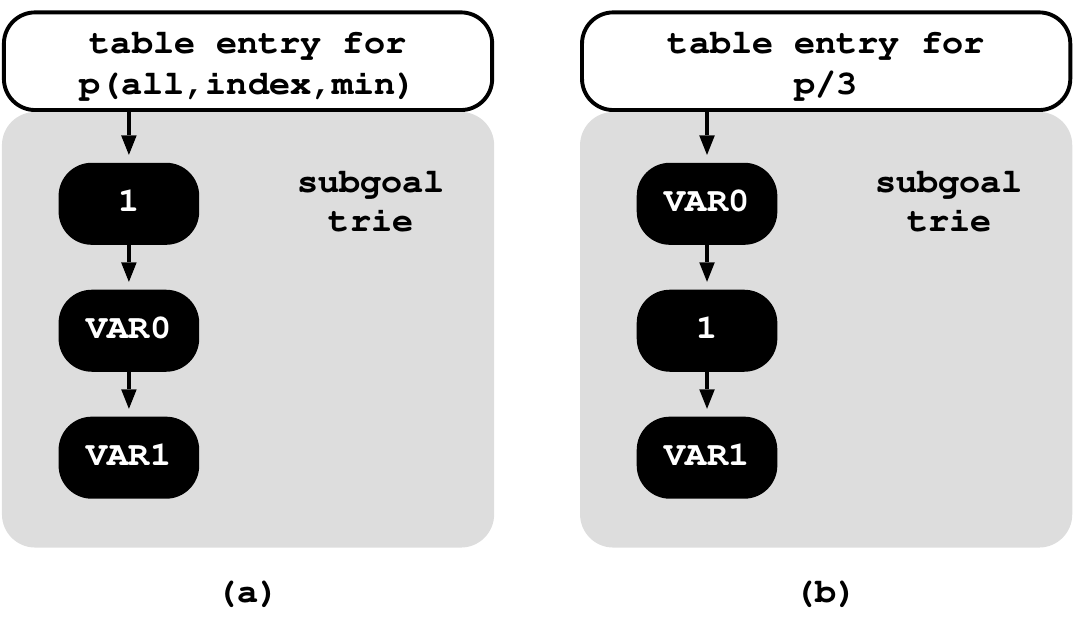}
\caption{Subgoal tries for \emph{p(X,1,Y)} considering \emph{p/3}
  declared (a) with and (b) without mode-directed tabling}
\label{fig_subgoal_trie}
\vspace{-\bigskipamount}
\end{wrapfigure}

During tabled evaluation, new tabled subgoal calls are inserted in
their own subgoal tries by following the order of the arguments in the
call. With mode-directed tabling, we follow the order defined in the
corresponding mode array. For example, consider again the
mode-directed tabled predicate \emph{p/3} as declared in
Fig.~\ref{fig_mode_array} and the subgoal call
\emph{p(X,1,Y)}. Figure~\ref{fig_subgoal_trie} shows the difference
between the resulting subgoal tries with and without mode-directed
tabling. The values in the mode array indicate that we should start by
inserting first the second argument of the subgoal call ($1$), then
the third argument ($Y$ or $VAR_0$) and last the first argument ($X$
or $VAR_1$).

The mode information is used when creating the subgoal frame
associated with the subgoal call at hand. With mode-directed tabling,
subgoal frames were extended to include a new array, named
\emph{substitution array}, where the mode information is stored,
together with the number of free variables associated with each
argument in the subgoal call. The argument's order is the same as in
the mode array.

\begin{wrapfigure}{r}{5.25cm}
\vspace{-\intextsep}
\centering
\includegraphics[width=5.25cm]{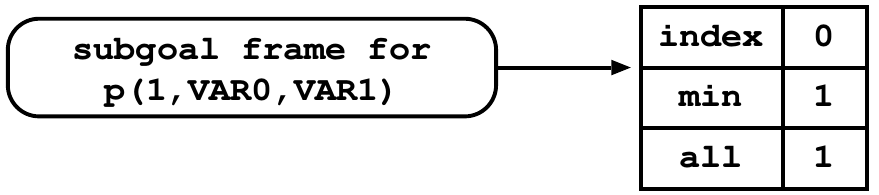}
\caption{Substitution array}
\label{fig_subs_array}
\vspace{-\bigskipamount}
\end{wrapfigure}

Figure~\ref{fig_subs_array} shows the substitution array for the
subgoal call \emph{p(X,1,Y)}. The first position, corresponding to the
argument with the constant 1, has no free variables and thus we store
a 0 in the substitution array. The other two arguments are free
variables and, thus, they have a 1 in the substitution array. It is
possible to optimize the array by removing entries that have 0
variables and by joining contiguous entries with the same mode. As we
will see next, the substitution array plays an important role in the
process of inserting answers in the answer trie.

\subsection{Mode-Directed Tabled Answers}
\label{sec_mode_directed_tabled_answers}

Like in traditional tabling, tabled answers are only represented by
the substitution terms for the free variables in the arguments of the
corresponding subgoal call. However, for mode-directed tabling, when
we are considering the substitution terms individually, it is
important to know beforehand which mode applies to each, and for that,
we use the information stored in the corresponding substitution
array. Moreover, the substitutions \emph{must be} considered in the
same order that the variables they substitute have been inserted in
the subgoal trie.

\begin{wrapfigure}{r}{6.35cm}
\vspace{-\intextsep}
\centering
\includegraphics[width=6.35cm]{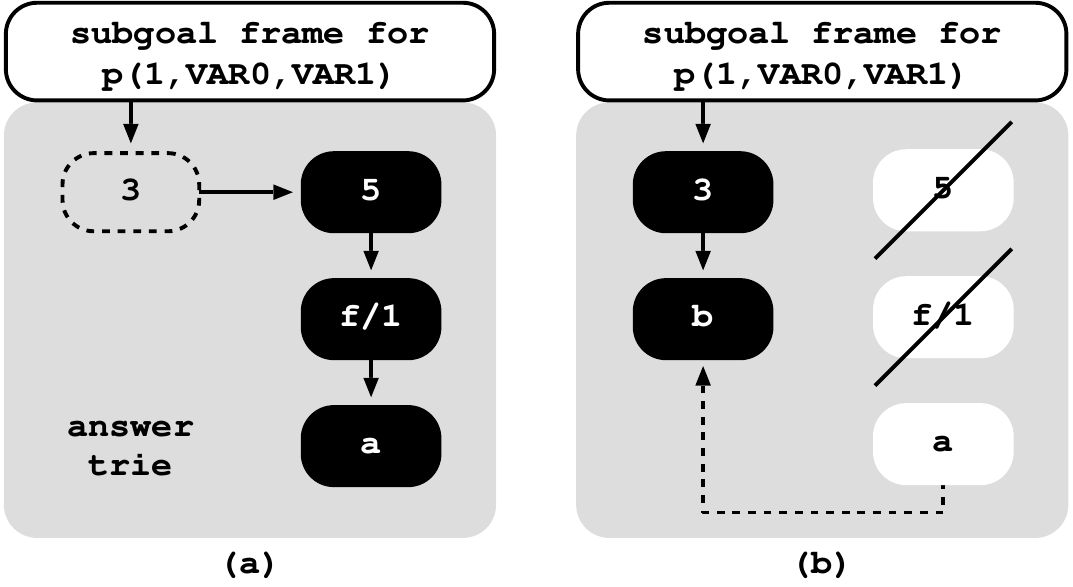}
\caption{Invalidating answers for \emph{p(X,1,Y)} (a) before and (b)
  after the invalidation process}
\label{fig_answer_trie_invalidation}
\vspace{-\intextsep}
\end{wrapfigure}

Consider again the substitution array for the subgoal call
\emph{p(X,1,Y)}. Now, if we find the answer \emph{\{X=f(a),~Y=5\}},
the first binding to be considered is \emph{\{Y=5\}} with \emph{min}
mode and then \emph{\{X=f(a)\}} with \emph{all} mode. Since the answer
trie is initially empty, both terms can be inserted as usual. Later,
if another answer is found, for example, \emph{\{X=b,~Y=3\}}, we begin
the insertion process by considering the binding \emph{\{Y=3\}} with
\emph{min} mode. As there is already an answer in the table, we must
compare both accordingly to the \emph{min} mode. Since the new answer
is preferable ($3<5$), the old answer must be \emph{invalidated} and
the new one inserted in the table. The invalidation process consists
in: (a) deleting all intermediate nodes corresponding to the answers
being invalidated; and (b) tagging the leaf nodes of such answers as
\emph{invalid nodes}. Invalid nodes are only deleted when the table is
later completed or
abolished. Figure~\ref{fig_answer_trie_invalidation} illustrates the
aspect of the answer trie before and after the invalidation process.

Invalid nodes are opaque to subsequent subgoal calls, but can be still
visible from the consumer calls already in evaluation. Hence, when
invalidating a node, we may have consumers still pointing to it. By
deleting leaf nodes, this would make consumers unable to follow the
chain of answers. An alternative would be to traverse the stacks and
update the consumers pointing to invalidated answers, but this could
be a very costly operation.

\begin{wrapfigure}{r}{6.35cm}
\vspace{-\intextsep}
\centering
\includegraphics[width=6.35cm]{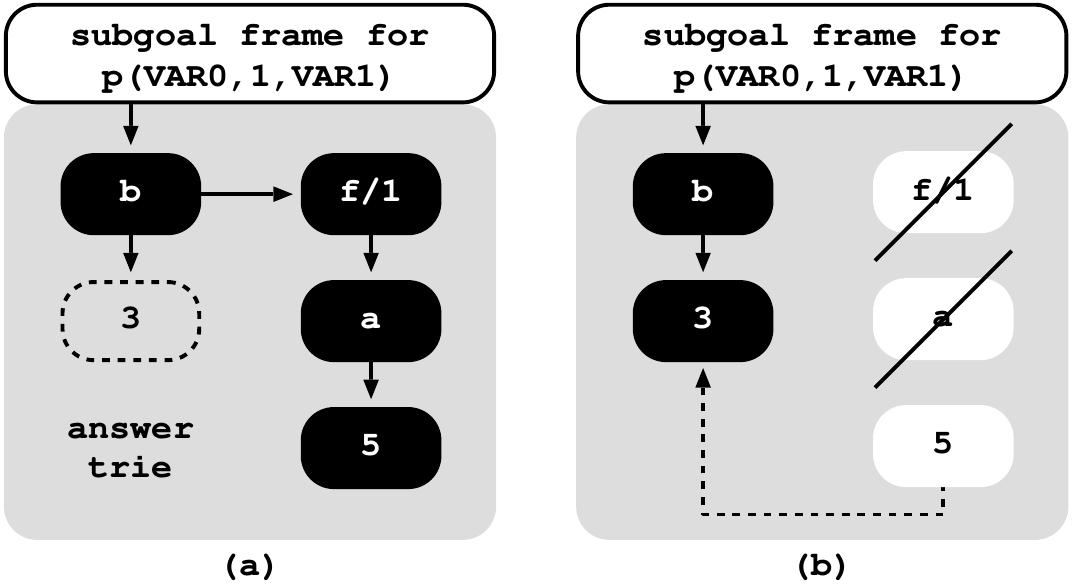}
\caption{Invalidating answers, without changing the insertion order,
  for \emph{p(X,1,Y)} (a) before and (b) after the invalidation
  process}
\label{fig_answer_trie_original_order}
\vspace{-\bigskipamount}
\end{wrapfigure}

Notice also that the mode's order in the substitution array is crucial
for the simplicity and efficiency of the invalidation process. When,
at a given node $N$, we decide that an answer should be invalidated,
the substitution array's order ensures that all nodes below node $N$
(including $N$) are the ones we want to invalidate and that the upper
nodes are the ones we want to keep. This might not be the case if we
used the original order. For example, consider again the call
\emph{p(X,1,Y)} and the answers \emph{\{X=f(a),~Y=5\}} and
\emph{\{X=b,~Y=3\}}. Figure~\ref{fig_answer_trie_original_order}
illustrates the invalidation process of these answers, if using the
original declaration.

To detect that the second answer is preferable ($3<5$), we need to
navigate in the trie until reaching the leaf node 5 for the first
answer. Thus, the invalidation process may require deleting upper
nodes (as the example in Fig.~\ref{fig_answer_trie_original_order}
shows) and/or traverse several paths to fully detect all preferable
answers (this would be the case if we had two intermediate answers
with the same minimal values, for instance \emph{\{X=f(a),~Y=5\}} and
\emph{\{X=h(c),~Y=5\}}), making therefore the invalidation process
much more complex and costly.


\subsection{Scheduling and Mode-Directed Tabling}

In a tabled evaluation, there are several points where we may have to
choose between continuing forward execution, backtracking, consuming
answers, or completing subgoals. The decision on which operation to
perform is determined by the scheduling strategy. The two most
successful strategies are \emph{batched scheduling} and \emph{local
  scheduling}~\cite{Freire-96}.

Batched scheduling evaluates programs in a depth-first manner as does
the WAM. When new answers are found for a particular tabled subgoal,
they are added to the table space and the evaluation continues with
forward execution. Only when all clauses have been resolved, the newly
found answers will be forwarded to the consumers. Batched scheduling
thus tries to delay the need to move around the search tree by
batching the consumption of answers.

Local scheduling is an alternative scheduling strategy that tries to
complete subgoals as soon as possible. The key idea is that whenever
new answers are found, they are added to the table space, as usual,
but execution fails. Local scheduling thus explores the whole search
space for a tabled predicate before returning answers for forward
execution.

To the best of our knowledge, YapTab is the only tabling system that
supports the dynamic mixed-strategy evaluation of batched and local
scheduling within the same evaluation~\cite{Rocha-05c}. This is very
important, because for mode-directed tabled predicates, the ability of
being able to use local evaluation can be crucial to correctly and/or
efficiently support some modes.

\begin{wrapfigure}{r}{6.5cm}
$:-~table~num\_links(index,sum).$\\
$num\_links(A,0):-~edge(\_,A).$\\
$num\_links(A,1):-~edge(A,\_).$\\

$:-~table~num\_nodes(sum).$\\
$num\_nodes(0).$\\
$num\_nodes(1):-~num\_links(\_,\_).$\\

$edge(a,b).~~~~~~edge(a,c).~~~~~~edge(b,c).$
\caption{A cascade of two mode-directed tabled predicates using the
  \emph{sum} mode}
\label{fig_sum_mode}
\vspace{-\bigskipamount}
\end{wrapfigure}

This is the case for the \emph{sum} mode. As it sums all the answers
for a given argument, we might end with wrong results if we return
partial results instead of aggregating them and only returning the
aggregated result. Consider, for example, the two mode-directed tabled
predicates $num\_links/2$ and $num\_nodes/1$ in
Fig.~\ref{fig_sum_mode} and the query goal $num\_nodes(N)$. If
$num\_links/2$ is evaluated using local scheduling, we get the right
result (\emph{N=3}) but, with batched scheduling, we end with a wrong
result (\emph{N=6}). This occurs because, with batched evaluation, the
$num\_links(\_,\_)$ call in the second clause of $num\_nodes/2$
succeeds 2 times for each $edge/2$ fact.

\begin{wrapfigure}{r}{6cm}
\vspace{-\intextsep}
\centering
\includegraphics[width=6cm]{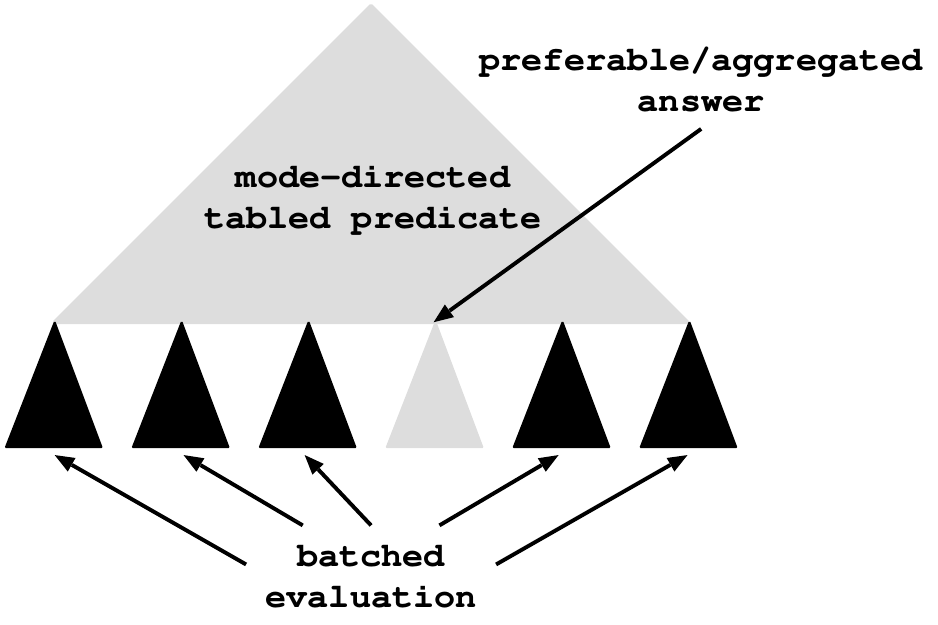}
\caption{Useless computations with batched evaluation}
\label{fig_scheduling}
\vspace{-2\bigskipamount}
\end{wrapfigure}

Batched evaluation can also yield useless computations for
mode-directed tabled predicates (see
Fig.~\ref{fig_scheduling}). Consider, for example, a mode-directed
tabled predicate $p/1$ declared as $p(max)$ and the query goal:

\vspace{0.25cm}
$:-~p(Max),~do\_work(Max,Res).$
\vspace{0.25cm}

With batched evaluation, the call to $do\_work(Max,Res)$ will be
executed for each $Max$ partial result computed by $p(Max)$, hence
originating as many useless computations as the number of non-maximal
results.


\section{Experimental Results}

In this section, we present some experimental results for a set of
benchmarks that take advantage of mode-directed tabling. The
environment for our experiment was a machine with a AMD FX(tm)-8150
8-core processor with 32 GBytes of main memory and running the Linux
kernel 64 bits version 3.2.0. To put our results in perspective, we
compare our implementation, on top of Yap Prolog (development version
6.3), with the B-Prolog (version 7.8 beta-6) and the XSB (version
3.3.6) systems, both using local scheduling. For XSB, we adapted the
benchmarks to use lattice answer subsumption (as discussed in
Section~\ref{sec_related_work})\footnote{For programs using
  \emph{min}/\emph{max} modes, we also tried with partial order answer
  subsumption but, unexpectedly, we got worst results.}. For
benchmarking, we used the following set of programs:

\begin{description}
\item [shortest(N)] uses the \emph{min} mode to determine all-pairs
  shortest paths in a graph representing the flight connections
  between the N busiest commercial airports in
  US\footnote{\url{http://toreopsahl.com/datasets}}.
\item [shortest\_first(N)] uses the \emph{first} mode to extend the
  all-pairs shortest paths program to also include the first
  justification for each shortest path.
\item [shortest\_all(N)] uses the \emph{all} mode to extend the
  all-pairs shortest paths program to also include all the
  justifications for each shortest path.
\item [shortest\_pref(N)] uses the \emph{last} mode to solve the 
  all-pairs shortest paths program using Preferences~\cite{Santos-12}.
\item [knapsack(N)] uses the \emph{max} mode to determine the maximum
  number of items to include in a collection, from N weighted items,
  so that the total weight is equal to a given value.
\item [lcs(N)] uses the \emph{max} mode to find the longest
  subsequence common to two different sequences of size N.
\item [matrix(N)] uses the \emph{min} mode to implement the matrix
  chain multiplication problem that determines the most efficient way
  to multiply a sequence of N matrices.
\item [pagerank(N)] uses the \emph{sum} mode to measure the rank
  values of web pages in a realistic dataset of web links called
  \emph{search
    engines}\footnote{\url{http://www.cs.toronto.edu/~tsap/experiments/download/download.html}},
  using N iterations.
\end{description}

Table~\ref{tab_times} shows the execution times, in milliseconds, for
running the benchmarks with YapTab, B-Prolog and XSB. In parentheses,
it also shows the overhead ratios against YapTab with local
evaluation. The execution times are the average of 3 runs. The entries
marked with $n.a.$ correspond to programs using modes not available in
B-Prolog. The ratios marked with (---) mean that we are \emph{not
  considering} them in the average results (they correspond either to
$n.a.$ entries or to execution times much higher than YapTab).

\begin{table}[t]
\centering
\caption{Execution times, in milliseconds, for YapTab, B-Prolog and
  XSB and the respective overhead ratios when compared with YapTab's
  local evaluation}
\begin{tabular}{l@{~~~}r@{~~~}|r@{~~~}r@{~~~}r}
\hline\hline
\multirow{2}{*}{\bf Programs}
& \multicolumn{2}{c}{\multirow{1}{*}{\bf YapTab}}
& \multicolumn{1}{c}{\multirow{2}{*}{\bf B-Prolog}} 
& \multicolumn{1}{c}{\multirow{2}{*}{\bf XSB}} \\
& \multicolumn{1}{c}{\multirow{1}{*}{\bf Local}} 
& \multicolumn{1}{c}{\multirow{1}{*}{\bf Batched}} \\
\hline
{\bf shortest(300)}        &  1,088 &  1,261 (1.16) &  2.990 (2.37) &     2,922 (2.69) \\
{\bf shortest(400)}        &  1,544 &  1,785 (1.16) &  4,216 (2.36) &     4,321 (2.80) \\
{\bf shortest(500)}        &  2,170 &  2,472 (1.14) &  5,792 (2.34) &     6,218 (2.87) \\
\hline
{\bf shortest\_first(300)} &  1,394 &  2,641 (1.89) &  3,225 (1.22) &     5,013 (3.60) \\
{\bf shortest\_first(400)} &  2,052 &  3,432 (1.67) &  4,614 (1.34) &     7,257 (3.54) \\
{\bf shortest\_first(500)} &  2,866 &  4,228 (1.57) &  7,400 (1.42) &    10,328 (3.60) \\
\hline
{\bf shortest\_all(300)}   &  4,324 &  8,383 (1.94) &  $n.a.$ (---) &     61,803 (---) \\
{\bf shortest\_all(400)}   &  5,861 & 10,590 (1.81) &  $n.a.$ (---) &    122,985 (---) \\
{\bf shortest\_all(500)}   &  8,337 & 13,598 (1.63) &  $n.a.$ (---) &    239,451 (---) \\
\hline
{\bf shortest\_pref(300)}  &  2,882 &  4,241 (1.47) &  $n.a.$ (---) &     6,666 (2.31) \\
{\bf shortest\_pref(400)}  &  4,152 &  5,621 (1.35) &  $n.a.$ (---) &     9,932 (2.39) \\
{\bf shortest\_pref(500)}  &  5,773 &  7,473 (1.29) &  $n.a.$ (---) &    14,129 (2.45) \\
\hline
{\bf knapsack(1000)}       &  1,013 &    998 (0.99) &    837 (0.84) &     2,684 (2.65) \\
{\bf knapsack(1500)}       &  1,581 &  1,561 (0.99) &  1,229 (0.79) &     3,977 (2.52) \\
{\bf knapsack(2000)}       &  2,037 &  2,040 (1.00) &  1,582 (0.78) &     5,473 (2.69) \\
\hline
{\bf lcs(1000)}            &  1,196 &  1,416 (0.98) &  2,900 (2.48) &     3,060 (2.56) \\
{\bf lcs(1500)}            &  2,768 &  3,560 (0.98) &  5,784 (2.12) &     7,128 (2.58) \\
{\bf lcs(2000)}            &  4,864 &  6,053 (0.99) & 10,116 (2.11) &    13,338 (2.74) \\
\hline
{\bf matrix(100)}          &    192 &    224 (1.17) &    582 (2.60) &       396 (2.06) \\
{\bf matrix(150)}          &    925 &  1,076 (1.16) &  2,549 (2.37) &     1,610 (1.74) \\
{\bf matrix(200)}          &  3,005 &  3,534 (1.18) &  7,816 (2.21) &     4,688 (1.56) \\
\hline
{\bf pagerank(1)}          &    365 &  $n.a.$ (---) &  $n.a.$ (---) &    128,377 (---) \\
{\bf pagerank(16)}         &    813 &  $n.a.$ (---) &  $n.a.$ (---) &  $>10~min$ (---) \\
{\bf pagerank(36)}         &  1,260 &  $n.a.$ (---) &  $n.a.$ (---) &  $>10~min$ (---) \\
\hline
\multicolumn{2}{l}{\it Average}     &        (1.29) &        (1.82) &           (2.49) \\
\hline\hline
\end{tabular}
\label{tab_times}
\vspace{-1.25\bigskipamount}
\end{table}

In general, the results show that, for all combinations of experiments
and systems, there is no clear tendency showing that the overhead
ratios increase or decrease as we increase the size of the
corresponding set of programs.

Comparing the results for local and batched evaluation, they show
that, on average, batched evaluation is around 29\% worse than local
evaluation. Batched evaluation gets worse the more answers are
inserted into the table space. This affects in particular the
\textbf{shortest\_first()}, \textbf{shortest\_all()} and
\textbf{shortest\_pref()} set of programs, which confirms our
discussion regarding the fact that batched evaluation is more suitable
to useless computations.

Regarding the comparison with the other systems, the results obtained
for YapTab clearly outperform those of B-Prolog and XSB. On average,
B-Prolog and XSB are, respectively, around 1.82 and 2.49 times worse
than YapTab using local evaluation.

Please note that for B-Prolog and XSB we do not include the
performance of some programs into the average results. For B-Prolog,
this is because these programs use the \emph{all}, \emph{last} and
\emph{sum} modes, which are not supported in B-Prolog. For XSB, the
execution times for the \textbf{shortest\_all()} and
\textbf{pagerank()} are much higher than YapTab and including them
would have distorted the comparison between the three systems. To the
best of our knowledge, YapTab is thus the only system that supports
the \emph{all}, \emph{last} and \emph{sum} modes and handles them
efficiently.


\section{Conclusions}

We discussed how we have extended and optimized YapTab's table space
organization to provide engine support for mode-directed tabling. In
particular, we presented how we deal with mode-directed tabled subgoal
calls and answers and we discussed the role of scheduling in
mode-directed tabled evaluations. Our implementation uses a more
general approach to the declaration and use of mode operators and,
currently, it supports 7 different modes. To the best of our
knowledge, no other tabling system supports all these modes and, in
particular, the \emph{sum} mode is not supported by any other
system. Experimental results on benchmarks that take advantage of
mode-directed tabling, showed that our implementation clearly
outperforms the B-Prolog and XSB state-of-the-art Prolog tabling
systems. In particular, YapTab is the only system that efficiently
handles programs that use the \emph{all} mode. Further work will
include extending our implementation to support multi-threaded
mode-directed tabling.


\section*{Acknowledgments}

This work is partially funded by the ERDF (European Regional
Development Fund) through the COMPETE Programme and by FCT (Portuguese
Foundation for Science and Technology) within projects PEst
(FCOMP-01-0124-FEDER-022701), HORUS (PTDC/EIA-EIA/100897/2008) and
LEAP (PTDC/EIA-CCO /112158/2009). João Santos is funded by the FCT
grant SFRH/BD/76307/2011.


\bibliographystyle{splncs}
\bibliography{references}

\end{document}